\documentstyle[12pt]{article}

         \def\la{\label}
         \def\rf{\ref}
         
         \def\n{\nu}
         \def\be{\begin{equation}}
         \def\bea{\begin{eqnarray}}
         \def\ee{\end{equation}}
         \def\eea{\end{eqnarray}}
         \def\se{\section}
         \def\sse{\subsection}
         \def\o{\over}

\begin{document}
\begin{titlepage}
\vspace{2mm}
\begin{center} {\Large \bf Some General Features of
                \\
\vskip 0.35cm
Matrix Product States in Stochastic Systems
\footnote{ to appear in Jour. Phys. A; Math. Gen.}}
\vskip 1cm
\centerline {\bf V. Karimipour
\footnote {e-mail:vahid@theory.ipm.ac.ir}}
\vskip 0.35cm
{\it Department of Physics, Sharif University of Technology, }\\
{\it P.O.Box 11365-9161, Tehran, Iran }\\
{\it Institute for Studies in Theoretical Physics and Mathematics,}\\
{\it P.O.Box 19395-5746, Tehran, Iran}\\
\vskip 2cm
\centerline{{\it Dedicated to Prof. Ardalan, my teacher, colleague, and friend,}}
\centerline{{\it on the occasion of his 60th birthday.}}

\end{center}

\vskip 2cm
\begin{abstract}
\noindent
We will prove certain general relations in Matrix Product Ansatz for one
dimensional stochastic systems, which are true in both random and sequential
updates. We will derive general MPA expressions for the currents and current
correlators and find the conditions in the MPA formalism, under which the
correlators are site-independent or completely vanishing. 
\end{abstract}

\end{titlepage}
\newpage
\section{Introduction}
One of the fruitful techniques for the study of stochastic systems on one
dimensional lattices is the Matrix Product Ansatz (MPA)\cite{hn}\cite{ks}, which is a
generalization of the simple product measure, where the steady state
probabilities are represented by matrix elements or traces of product of
appropriate operators. This ansatz when applied to processes with random
sequential updates, states that the steady state of a process governed by a
Hamiltonian of the form
\be \la{4} H = h_1 + \sum_{i=1}^{N-1} h^B_{k,k+1} + h_N. \ee
can be written as:
\be \la{5} |P> = {1\o Z_N} <W|{\cal A }\otimes{\cal A }\otimes
\cdots {\cal A }|V >. \ee where $Z_N$ is a normalization constant.
Here $ {\cal A }$ is a column matrix with operator entries acting
on some auxiliary space  $ F $, that is
\be \la{6} {\cal A } = \sum_{i=0}^p A_i |i> = \left( \begin{array}
{l}A_0\\ A_1 \\ A_2\\. \\ .\\ A_p\end{array}\right) \ee
and $ < W | $ and $ | V > $ are two vectors in $ F^* $ and $ F $ respectively.
Here we have assumed that the Hilbert space of states of the chain is
$ {\cal H}={\bf h}^{\otimes N} $, where ${\bf h}$ is the $p+1$ dimensional
space of one site. The space ${\bf h}$ is spanned by the vectors $|i> ; i=0,\cdots p $
where $|0> $ stands for a vacant site and $|i>$ stands for a site occupied by a
particle of type $i$.
Note that we use the same symbol $ | > $ for a vector in ${\bf h}$
and $ F $, hoping that
no confusion will arise.\\ The conditions for stationarity  of (\rf{5})
are the following \cite{ks}
\be\la{7} h^B {\cal A}\otimes {\cal A} = {\cal X} \otimes {\cal A} -
{\cal A}\otimes {\cal X} \ee
\be\la{8} (h_N{\cal A} - {\cal X}) \vert V> = 0 \ee
\be\la{9} <W\vert (h_1{\cal A} + {\cal X}) =0\ee
where $ {\cal X} $ is a suitably chosen vector with in general operator
entries, i.e:
\be \la{a} {\cal X } = \sum_{i=0}^p X_i |i> = \left( \begin{array}
{l}X_0\\ X_1 \\ X_2\\. \\ .\\ X_p\end{array}\right) \ee
When applied to processes with backward sequential (BS) updates
in discrete time with updating operator
\be \la{12} {\cal T } = T_1 T^B_{12}T^B_{23}\cdots T^B_{N-1 N} T_N\ee
where $ T_1 $ and $ T_N $ are the boundary terms and the rest
of operators implement the bulk dynamics,
the steady state $ |P>$ is written in the same form as in (\rf{5}),
but the MPA realtions (\rf{7}-\rf{9}) are replaced by [\cite{h},\cite{rss}]
\be \la{13} T^B {\cal A}\otimes {\hat {\cal A}}  = {\hat {\cal A}}
\otimes {\cal A} \ee \be \la{14} T_N {\cal A}|V> = {\hat {\cal A}}|V>\ee
\be \la{15} <W|T_1 {\hat {\cal A}} = <W|{\cal A}\ee
where $ {\hat A}$ is a new operator-valued vector in ${\bf h}$.
\be \la{16} {\hat {\cal A }} = \sum_{i=0}^p {\hat A}_i |i> = \left(
\begin{array}{l}{\hat A}_0\\ {\hat A}_1 \\ {\hat A}_2\\. \\ .\\ {\hat A}_p
\end{array}\right) \ee
The form of the relations for forward sequential (FS) update is obtained from
(\rf{13}-\rf{15}) by  interchange of ${\cal A} $ and $ {\hat {\cal A}}$.\\
This ansatz has been applied to the Asymmetric Simple Exclusion Process
(ASEP) under
various circumstances and with different kinds of updates ( see \cite{d}
,\cite{sch} and \cite {de} and references therein.) To explore the
usefulness of MPA beyond the above simple cases, there have also been a number of attemps to study the
algebras associated with more complicated processes \cite{ev} \cite{evg}\cite{rmm}\cite{lpk} and \cite{m}, or conversely, the
processes associated with more complicated algebras\cite{ahr}\cite{adr}\cite{r60}\cite{k1} and \cite{k2}.
In pursuing the latter line, that is begining from algebras and searching for
processes, one usually notices that the kind of algebra severely restricts the
kind and the rates of the processes,  especially on open systems. \\
The aim of this paper is to start from the general MPA algebras (\rf{7}-\rf{9}) and (\rf{13}-\rf{15}) and
derive some general relations, which are valid for a large class of
processes. These relations put some general constraints on the algebra that one
wants to start with for the formulation of a process in the MPA formalism. We
first state our main results the proof of which will be found in the
forthcoming sections.\\  \\
{\large\bf Main Results}:\\
We consider the family of processes on an open chain, described by Hamiltonians
of the type (\rf{4}),
where there are a number of species of particles interacting in the bulk. The
processes in the bulk are quite arbitrary, i.e. particles can be created,
annihilated or can coagulate or decoagulate. However we assume that some species
of particles are conserved, so that for each conserved species say the
$i$-th one, a conserved current $ J^i$ can be defined. Particles are injected
to
or extracted from both the left and the right ends and a driving force may also
be present. \\
Due to this arbitrariness we do not use the detailed form of any particular
Hamiltonian or the associated algebra, or any of its representations therof, but
only use the basic formulas of MPA.
Our main results are the following:\\
{\bf 1-}If a species say the $i$-th one is conserved, then its current is given by:
$$ <J^i> = {<W|X_iC^{N-1}|V>\o <W|C^N|V>}$$
where in the  ordered updates $ X_i = A_i - {\hat A}_i$.\\
{\bf 2-}If $\sum_{i}X_i = 0 $ , then the $n$-point correlators of all the
conserved currents are independent of the distances between the points. This is
true even in finite open chains and is independent of what the other species are doing.\\
{\bf 3-}If a set of $\{X_{i_1},X_{i_2},\cdots\}$ are c-numbers, then the
connected correlation functions for the currents of particles
$\{i_1, i_2, \cdots\}$ vanish in the thermodynamic limit. Moreover in a finite
chain with N sites, one has for the above set of particles:
\be \la{m} <J^{r} J^{s}>_N = <J^s J^r>_N = <J^r>_N <J^s>_{N-1}\ee.
\be \la{mm} <J^{r} J^{s} J^t>_N = <J^r>_N <J^s>_{N-1}<J^t>_{N-2}\ee
etc.\\
{\bf Remarks:}\\
{\bf a:}The above properties when supplemented with the results from numerical computations
or simulations for small chains, may give us an idea of what kind of algebra can or
can not be used for an exact solution of the problem in a large chain.\\
{\bf b:}The above results are particular to open systems and do not apply
to processes on closed systems.\\  \\
{\bf 4-}Under any circumstances the currents and densities of the conserved
species in backward and forward sequential updates are related as follows:
$$ <n^i>_{\rightarrow}(k) - <n^i>_{\leftarrow} (k) =
<{J^i}_{\rightarrow}> = <{J^i}_{\leftarrow}> $$
where the direction of the arrows indicate the type of update. This result is a generalization
of a previous one first derived in \cite{rs}.\\
{\bf Remark:} In deriving the above results we have not used any particular
representation of the MPA algebra, i.e: like the Fock space representations
of \cite{rs} or \cite{ks}.\\

\se{ General relations in MPA algebras }

In order to deal with all the different updates in a uniform manner we define
the following objects
\be \la{19} h_1:= 1-T_1, \hskip 1cm h_N:= 1-T_N, \hskip 1cm h^B := 1-T^B \ee
and define in the BS and FS updates the vector
\be \la{20} {\cal X}:= {\cal A}- {\hat {\cal A}} .\ee

With these defintions, the MPA relations for these ordered updates take the
following form, compared with those for the random update (see eqs. (1-4)).
\be\la{24} h^B {\cal A}\otimes ({\cal A}-{\cal X}) = {\cal X} \otimes
{\cal A} - {\cal A}\otimes {\cal X} \ee
\be\la{25} (h^N{\cal A} - {\cal X}) \vert V> = 0 \ee
\be\la{26} <W| (h^1{\hat {\cal A}} + {\cal X}) =0\ee
Note the similarity between the equations for different updates.
In fact (\rf{8}) and (\rf{25}) are exactly the same. This rewriting
could be of no use were it not for the fact that usually (see for example
\cite{rsss}\cite{fj}), the evolution operator $ T^B$ of a process in ordered update
is, modulo a redefinition of parameters, nothing but $ 1+h^B$ where $h^B$ is
the Hamiltonian of the same process in random sequential update. Therefore this
rewriting
allows one to immediately write down the MPA relations
for any of these updates, once they are known for one of them.
Furthermore, in those situations where a map may exist between the MPA algebras
for different updates, such rewriting facilitates the search for such a map.
We will see an example of this
later on.
Before proceeding further we
fix some notations and conventions. In the Hilbert space ${\bf h} $ we define
a reference state $ <s|$ as
\be\la{27} <s| := \sum_{i=0}^{i=p}<i| \equiv ( 1,1,\cdots 1)\ee
The reference state for the space $ {\bf h}\otimes {\bf h} $ is defined as
$ <ss|:= <s|\otimes <s| $
and similarly for tensor products having more factors.
This state is used to write the sum of entries in a column of a vector or a
matrix in closed form.
The following operators acting on the space $ F $ are also useful:
\be \la{28} C:= \sum_{i=0}^{i=p} A_i = <s|{\cal A}> \hskip 2cm K:=
\sum_{i=0}^{i=p} X_i = <s|{\cal X}>\ee
The normalization constant $ Z_N$ is given by $ Z_N = <W|C^N|V>$.
Conservation  of probabilities imply that we should have in all three updates
\be \la{29} <ss| h^B = <s|h_1 = <s|h_N = 0 \ee
Actually in ordered sequential updates, conservation of probability
need not hold
in every single update, but only in one complete update. Therefore one
should only
have $ <S| {\cal T} = <S| $. In writing (\rf{29}) we are assuming
conservation of probability
at each single update, i.e. $ <ss| T^B = <ss|$. The known sequential
updating procedures for the ASEP fall within this class.\\
Multiplying both sides of the MPA relations (\rf{7} and \rf{8}) or
(\rf{24} and \rf{25}) ({from the left by $ <ss|$ or $ <s| $ where
appropriate and using (\rf{28}) and (\rf{29}) we obtain the
following relations which are valid in all kinds of updates
\be \la{30} K|V> = <W|K = 0 \hskip 2cm [ K , C] = 0.\ee
This relation has been obtained first in \cite{rs}.\\
We now consider a processes in which some of
the particle species are conserved locally.
In this case the local operator $ h^B_{k,k+1}$ does
not change the number of particles in a conserved species say the $i-$th one, on the couple of
sites $ k$ and $ k+1 $.
Thus we have
\be \la{32} [ {\hat \tau}^i_k + {\hat \tau}^i_{k+1} , h^B_{k,k+1} ] = 0 \ee
where $ {\hat \tau}^i$ is the number operator of particles of type $i$, i.e.
${\hat \tau}^i |j> = \delta_{ij}|j>$.
Applying the state $ <ss|\tau^i_k + \tau^i_{k+1} $ on both sides of
(\rf{7} or \rf{24}) and using (\rf{28} and \rf{29}) we obtain:
\be \la{33} [ X_i , C ] = [ A_i , K]  \ \ \ \ \ \ \ \ \ \ \ \forall i \ee
This relation holds for each conserved species seperately and is a new
constraint on the MPA algebra for such processes.\\
{\bf Remark:} In a process consisting only of exchange of particles, more
detailed relations can be obtained. However eq. (\rf{33}) is valid independent of
the interaction of the other species.
\se{The MPA expression for the currents}
The average density of particles of type $i $ at site $ k $ is defined as
$ <\tau^i_k>(t) := <S|{\hat \tau}^i_k |P(t)>$
where $ <S|$ is the reference state of the whole lattice.
We will now obtain general MPA expressions for currents
of conserved species of particles.\\
At least in the RS update, the MPA expressions for the current of a conserved
species can be derived simply by
calculating the current at one of the boundary links, say the rightmost
link. This current is
the product of density at the rightmost site and the extraction rate of
that species. However we  prefer to follow a different approch and calculate
the currents directly in the bulk, in order to calculate also the current
correlators (see section {\bf 3.1}). \\
{\bf 1) The RS  update:}\\
According to the Hamiltonian formulation of Markov processes, we
have
\be \la{41} {d\o dt} <\tau^i_k> = <S|[ H , {\hat \tau}^i_k ]|P(t)>.\ee
Due to the form of $H$ this can be rewritten as
\be \la{42} {d\o dt} <\tau^i_k> = <S|[ h^B_{k-1,k}+h^B_{k,k+1} ,
{\hat \tau}^i_k ]|P(t)>.\ee
Combination of (\rf{32}) and (\rf{42}) now yields
\be \la{43} {d\o dt} <\tau^i_k> = <S|[ h^B_{k-1,k},
{\hat \tau}^i_k ]|P(t)> - <S|[ h^B_{k,k+1} , {\hat \tau}^i_{k+1}]|P(t)>\ee
which can be written as a continuity equation:
\be \la{44} {d\o dt} <\tau^i_k> = <J^i_k>-<J^i_{k+1}>. \ee
with the current of particles of type $i$ into site $k$ being
\be \la{45} <J^i_k> = < S|[h^B_{k-1,k}, {\hat \tau}^i_k]|P(t)>\ee
To find the MPA relation for the steady state currents we rewrite (\rf{45}) as
\be \la{46} <J^i_k> = <W|{\hat J}^i_{k}|V> \ee
where ${\hat J}^i_{k}$ is the following operator acting on $ F$
\be \la{47} {\hat J}^i_{k} = {1\o Z_N}< S|[h^B_{k-1,k}, {\hat \tau}^i_k]
|{\cal A }\otimes {\cal A }\otimes \cdots {\cal A }>.\ee
Note that here we are looking at $|{\cal A }\otimes
{\cal A }\otimes \cdots {\cal A }>$
as an operator valued vector in ${\cal H}$.
Using (\rf{28}), we find
\be \la{48} {\hat J}^i_{k} = {1\o Z_N} C^{k-2}<ss|[h^B , 1\otimes \tau^i]
|{\cal A }\otimes {\cal A }> C^{N-k} =: {1\o Z_N}C^{k-2} M_i C^{N-k}\ee
The operator $M_i$ is calculated as follows:
\bea \la{49} M_i &=& <ss| h^B (1\otimes {\hat \tau}^i)- (1\otimes
{\hat \tau}^i) h^B |{\cal A }\otimes {\cal A }>\cr
&=& -<ss|1\otimes {\hat \tau^i}|{\cal X} \otimes {\cal A} -
{\cal A}\otimes {\cal X} >\cr
&=& -KA_i + CX_i \eea
where in the second line we have used (\rf{7} and \rf{29}) and in
the third we have used
(\rf{6}),(\rf{a}) and (\rf{28}).
Thus using (\rf{30}), we find:
\be \la{50} <J^i> = {<W|C^{k-2}X_iC^{N-k}|V>\o <W|C^N|V>}\ee
We can now use (\rf {33}) and rewrite $CX_i = X_i C + KA_i - A_iK$ and move the two
$K'$s to the left and right where their action on the vectors $ <W|$ and $|W>$
vanish. In this way we can move $ X_i$ completely to the left and obtain:
\be \la{50'} <J^i> = {<W|X_iC^{N-1}|V>\o <W|C^N|V>}\ee
This kind of relation for conserved currents has already been obtained for specific
2-species algebras\cite{adr} where all the particles are conserved. Here we derive it in quite general form.\\
{\bf 2) The BS and FS updates:}\\
In the
BS update we have:
\be \la{51} <\tau^i_k>(t+1) -<\tau^i_k>(t) = <S|{\hat \tau}^i_k{\cal T}|P(t)>
 - <S|{\hat \tau}^i_k|P(t)> \ee
Using the property $ <S|{\cal T} = <S|$ , we can write (\rf{51} ) in the form:
\be \la{52} <\tau^i_k>(t+1) -<\tau^i_k>(t) = <S|[\tau^i_k,{\cal T}]|P(t)>\ee
Taking the structure of $ {\cal T } $ into account (see (\rf{12})),
using the notation
$ {\cal T}^B_{k,l}:= T^B_{k,k+1}\cdots T^B_{l,l+1}$
( with $ T^B_{N,N+1}:=T_N$ )
and the property $<S|{\cal T}^B_{k,l} = <S|$, we can rewrite eq.(\rf{52}) as:
\bea \la{53} <\tau^i_k>(t+1) -<\tau^i_k>(t) &=& <S|[{\hat \tau}^i_k
,T^B_{k-1,k}T^B_{k,k+1}]
{\cal T}^B_{k+1,N}|P(t)> \cr &=& <S|[{\hat \tau}^i_k,T^B_{k-1,k}]
{\cal T}^B_{k,N}|P(t)>\cr &+&
 <S|[{\hat \tau}^i_k,T^B_{k,k+1}]{\cal T}^B_{k+1,N}|P(t)>\eea
where in the last line we have used $ <ss|T_{k-1,k}^B = <ss|$.
Using the local conservation law
\be \la{54} [{\hat \tau}^i_k + {\hat \tau}^i_{k+1}, T^B_{k,k+1}] = 0 \ee
Equation (\rf{53}) can be written in the form of a continuity equation
\be \la{55} <\tau^i_k>(t+1) -<\tau^i_k>(t) = <J^i_k>(t) -<J^i_{k+1}>(t)\ee
Where $ J^i_k $ is the number of $i$-particles which, in the interval
between $ t$ and $ t+1 $, leave site $k-1$ and enter into site $k$,
and is given by
\be \la{56} <J^i_k> = <S|[{\hat \tau}^i_k,T^B_{k-1,k}]{\cal T}^B_{k,N}|P(t)>\ee
The MPA relation for the current $ <J^i_k> $ is now obtained along the
same lines as in the RS update, namely
\be \la{57}<J^i_k> = <W| {\hat J}^i_k |V> \ee
where
\be \la{58} {\hat J}^i_{k} = < S|[{\hat \tau}^i_k, T^B_{k-1,k}]
{\cal T}^B_{k,N}|{\cal A }\otimes {\cal A }\otimes \cdots {\cal A }>.\ee
Using (\rf{13}) and (\rf{28}) we obtain
\bea \la{59} {\hat J}^i_{k} &=& < S|[{\hat \tau}^i_k, T^B_{k-1,k}]
|{\cal A }\otimes {\cal A }\otimes \cdots \underbrace {{\cal A}
\otimes {\hat {\cal A}}}_{k-1,k} \cdots {\cal A }>\cr
&=& C^{k-2}<ss|[1\otimes {\hat \tau}^i, T^B]|{\cal A }\otimes
{\hat {\cal A}}> C^{N-k} \cr
&=:& C^{k-2} M_i C^{N-k}\eea
The \ \ operator\ \  $ M_i $ \ \ is calculated \ \ by expanding\ \
the commutator\ \  and using
\ \  (\rf{13}) and \ \ \ \ \ $ <ss|T^B = <ss|$, with the result
\bea \la{60} M_i &=& <ss|1\otimes {\hat \tau}^i |{\hat {\cal A }}\otimes
{\cal A}> - <ss|1\otimes {\hat \tau}^i|{\cal A }\otimes {\hat {\cal A}}>\cr
&=& {\hat C}A_i - C {\hat A}_i \eea
Using the fact that $ {\hat C} = C - K, {\hat A}_i = A_i - X_i, $
we rewrite $ M_i $ as $ CX_i - KA_i $. Using the same argument as we did
in the RS case we find
\be \la{61} < {J^i}_{\leftarrow}> = {<W|X_iC^{N-1}|V>\o <W|C^N|V>}\ee
Similar manipulation shows that in the FS update the current is given by
the same general form as in (\rf{61}),
with $ C $ replaced with ${\hat C}$.
Writing $ {\hat C} = C+K $, expanding $ (C+K)^{(N-1)} $ and using $ K|V>=0$,
we find that
the currents of each conserved species in
these two updates are equal for arbitrary transition
probabilities. i.e.
\be \la{62} <{J^i}_{\rightarrow}> = <{J^i}_{\leftarrow}> =
{<W|X_iC^{N-1}|V>\o <W|C^N|V>}\ee
We have proved this relation under a more general condition as
in \cite{rs}, thats
we do not assmue $ C={\hat C}$ or equivalently $ K=0$. As we have pointed
out in the introduction and will prove in the next subsection
the condition $ K=0$ puts physical restrictions on the steady state, namely in
this case all the current correlators become distance-indepdent.
Moreover we find the following relation between density profiles and currents
in the two updates, which is valid regardless of the bulk and boundary
transition rates:
\be \la{63} <n^i>_{\rightarrow}(k) - <n^i>_{\leftarrow} (k) =
<{J^i}_{\rightarrow}> = <{J^i}_{\leftarrow}> \ee
the proof of which is easy to see, once we note that
\bea \la{64} <n^i>_{\rightarrow}(k) &=& {1\o {\hat Z}_N} <W|{\hat C}^{k-1}
{\hat A}_i {\hat C}^{N-k}|V>\cr
 &=& {1\o Z_N} <W|C^{k-1} (A_i-X_i) C^{N-k}|V>\cr &=& <n^i>_{\leftarrow}(k)
- {J^i}_{\rightarrow}\eea
This is the generalization of a similar relation for
for 1-ASEP \cite{rs}, which is now valid for each species seperately and for
arbitrary transition probabilities.
\sse{Equal time current correlators}
In this section we consider only the RS update and
find the MPA expressions for the current correlators
$<J^i_k J^j_l>$ of two kinds of particles $ i $ and $ j $ at sites k and l.
For $ l>k+1 $ (i.e. disjoint links) we obtain a simple relation.
Starting from the definition
\be <J^i_k J^j_l>:= <S|[h^B_{k-1,k}, {\hat \tau}^i_k] [h^B_{l-1,l},
{\hat \tau}^i_l] |P(t)>\ee
and proceeding exactly along the lines which led to (\rf{50}) we find
\be \la{*} <J^i_k J^j_l>_N = {<W|C^{k-1}X_iC^{l-k-2}X_jC^{N-l+1}|V>\o
<W|C^N|V>}\ee
The proof of this relation is detailed in the appendix.
For $l=k+1$ (i.e. consecutive links) no simple relation is obtained.
Several remarks are in order
now.\\
{\bf Remarks}:\\
{\bf 1:} In the special case $ K = 0$, the two point correlator
(and in fact all the
n-point current correlators (see the appendix)), become
site-independent, since in this case
the operators $ X_i $ commute with $ C$ and the two point
correlator can be written
as
\be \la{**} <J^i_k J^j_l>_N = {<W|X_i X_jC^{N-2}|V>\o <W|C^N|V>}\ee
{\bf 2:} For number-valued $ X_i $ and $ X_j$ this relation implies
\be \la{***} <J^i_k J^j_l>_N = <J^j_k J^i_l>_N = <J^i>_N <J^j>_{N-1}. \ee
with similar relations for higher correlators.
This is of course a finite-size effect and in the thermodynamic limit,
the currents at disjoint links are not correlated. However this is true only
when some of the $ X_i$'s are number-valued . In this way we have shown that the nature of
${\cal X}$, controlls the current correlators in a very definite way, namely
for general ${\cal X}$, but with $K:=\sum X_i = 0$, the correlators are site
independent and for number valued ${\cal X}$, there is no correlation in the
thermodynamic limit. \\
{\bf 3:} No such simple relations can be obtained for the FS and BS updates
even for disjoint links. This is due to that fact in these updates, no matter
how far are the links, the current operators for these links contain a
common string of local operators (see (\rf {56})).
Thus although the currents of the two ordered updates are equal, their correlators
may not be related to each other in any simple way. In this sense the steady state
of the two updates may not be physically equivalent.\\
\se { Beyond nearest neighbour interactions}
The matrix product ansatz as formulated in \cite{ks}, can be generalized
to models with
more general Hamiltonians.\footnote {The possibility of extending
MPA to include more non-local interactions has also been recently noted
in \cite{kls}}. In the following we consider a Hamiltonian with nearest
and next-nearest neighbor interactions, although our analysis can be
generalized to
more non-local Hamiltonians. Consider a Hamiltonian of the form:
\be \la{73} H = h_{12} + \sum_{k=1}^{N-2} h^B_{k,k+2} + h_{N-1,N}. \ee
where $ h^B_{k,k+2}$ acts on the three sites $ k,k+1$ and $ k+2$ and
$ h_{12}$ and $ h_{N-1,N}$ are boundary
terms. Writing the steady state as in (\rf{5}) one finds the
following conditions for
stationarity:
\be\la{74} h^B {\cal A}\otimes {\cal A}\otimes {\cal A} =
{\cal X} \otimes {\cal A} - {\cal A}\otimes {\cal X} \ee
\be\la{75} (h_{N-1,N}{\cal A}\otimes {\cal A} - {\cal X}) \vert V> = 0 \ee
\be\la{76} <W\vert (h_{12}{\cal A}\otimes {\cal A} + {\cal X}) =0\ee
where $ {\cal X}$ is in general an operator valued tensor in
$ {\bf h}\otimes {\bf h}$, i.e.
\be \la{77}{\cal X} := \sum_{\alpha, \beta} X_{\alpha, \beta}
| \alpha, \beta > \ee
and $ |\alpha> $ and $|\beta > $ stand for the states of one site.
Denoting as before $ C:= <s|{\cal A}>$ and $ K:= <ss|{\cal X}>$ one finds
again that equation (\rf{30}) is true
also in this case. To find the MPA expressions for the currents
we proceed as in
section 4 and find:
\be \la{78} {d\o dt} <\tau^i_k> = <S|[ h^B_{k-2,k} + h^B_{k-1,k+1} +
h^B_{k,k+2},{\hat \tau}^i_k ]|P(t)>\ee
Local conservation of particles, now implies
\be \la{79} [ h^B_{k-1,k+1}, {\hat \tau}^i_k ] = - [ h^B_{k-1,k+1}
 , {\hat \tau}^i_{k-1} + {\hat \tau}^i_{k+1}]\ee
Acting on (\rf{74}) by $<sss|{\hat \tau}^i_k\otimes 1 \otimes
1 + 1\otimes {\hat \tau}^i_k\otimes 1 + 1\otimes 1\otimes {\hat \tau}^i_k $
and using (\rf{28},\rf{29}) we find that
\be \la{80}[ X_i^{(1)} + X_i^{(2)} , C] = 0, \ee
where
\be \la{81}X_i^{(1)}:= \sum_j X_{ij} \hskip 2cm  X_i^{(2)} := \sum_j X_{ji}.\ee
Inserting (\rf{79}) in (\rf{78}), we find again a continuity equation as in
(\rf{44}) with the currents
\be \la{82}\la{36} <J^i_k> = <S|[ h^B_{k-2,k},{\hat \tau}^i_k ] -
[ h^B_{k-1,k+1},{\hat \tau}^i_{k-1}]|P(t)>\ee
The MPA expression for this current is obtained along the same
line as in section
4. After similar manipulations we find
\bea \la{83} {\hat J}^i_{k} &=&{1\o Z_N} C^{k-3}<sss|[h^B , 1\otimes 1\otimes
{\hat \tau}^i] |{\cal A }\otimes {\cal A }\otimes {\cal A}> C^{N-k} \cr
&-& {1\o Z_N} C^{k-2}<sss|[h^B , {\hat \tau}^i\otimes 1\otimes 1] |{\cal A }
\otimes {\cal A }\otimes {\cal A}> C^{N-k+1}\eea
Expanding the commutators and using (\rf{28}) , (\rf{29}) and (\rf{74}) we find
\be \la{84} {\hat J}^i_{k} = C^{k-2}<ss|(1\otimes \tau^i + \tau^i\otimes 1)
|{\cal X }> C^{N-k} \ee
which finally yields
\be \la{85}<{J}^i> = {<W|X^{(i)}C^{N-1}|V>\o <W|C^N|V>}\ee
where
\be \la{86}X^{(i)} := \sum_{\n}(X_{i,\n} + X_{\n,i}) \ee
The generalization of these results to Hamiltonians with more non-local
interactions is obvious.
\se{Discussion}
We have discussed the general structure of MPA states in stochastic systems
and have proved certain general relations for general stochastic
processes in random and ordered updates. We have tried to be as general as possible, our results
in sections 3-5 are independent of the bulk and boundary transition rates
and are also independent of the asymmetry caused by driving.
We have found general MPA expressions for the currents and current correlators
and have shown that when in the MPA formalism one uses
general operator-valued ${\cal X}$ but with $ \sum_i X_i = 0 $, the
current correlators while non-vanishing, become site independent.\\
And when any of the $X_i$'s are c-numbers, then the connected correlation
functions of the corresponding currents vanish in the thermodynamic limit.\\

\se{Acknowledgement}
I would like to thank V. Rittenberg, H. Hinrichsen and G. M. Sch{\"u}tz for very
valuble comments through email correspondnecs.
\se{Appendix}
In this appendix we present the proof of ({65}). In the same spirit that we
have derived the expressions for the currents we write
\be <J^i_k J^j_l> = <W|{\hat J}^i_k {\hat J}^j_l|V> \ee
where
\bea {\la g}{\hat J}^i_k {\hat J}^j_l \cr &=&{1\o Z_N}
<S|[h^B_{k-1,k} , {\hat\tau}^i_k][h^B_{l-1,l} , {\hat\tau}^i_l]|
{\cal A }\otimes {\cal A } \cdots {\cal A}>\cr &=& {1\o Z_N}C^{k-2}
M_i C^{l-k-2}M_j C^{N-l}\cr
&=&{1\o Z_N}C^{k-2}(CX_i - KA_i) C^{l-k-2}(X_jC - A_jK)C^{N-l}\eea
where in the last line we have used two equivalent expressions
for $ M $, in order to move the two $ K $'s to the left and right respectively
and act by them on $<W| $ and $|V>$ to obtain
\be <J^i_k J^j_l> = {<W|C^{k-1} X_i C^{l-k-2}X_jC^{N-l+1}|V>\o <W|C^N|V>}\ee
This result is true for the two point correlators, since in higher correlators
one can not eliminate all the $K$'s in (\rf {g}). When $ K=0$, this further
simplifies
to (\rf{**}). Furthermore in this case formula (\rf{**}) trivialy generalizes to n-point
functions.

\end{document}